\documentclass[%
 reprint,
 amsmath,amssymb,
 aps,
prl,
]{revtex4-1}

\usepackage{color}
\usepackage{graphicx}
\usepackage{dcolumn}
\usepackage{bm}

\begin{document}

\title{Spiral-based phononic plates: From wave beaming to topological insulators}

\author{Andr\'e Foehr,$^{1, 2}$ Osama R. Bilal,$^{2, 3*}$ Sebastian D. Huber,$^{3}$ and Chiara Daraio$^{2}$}
\email{bilalo@ethz.ch and daraio@caltech.edu}
\affiliation{${^1}$Department of Mechanical and Process engineering, ETH Zurich, 8092 Zurich, Switzerland}

\affiliation{${^2}$Division of Engineering and Applied Science, California Institute of Technology, Pasadena, California 91125, USA}
\affiliation{${^3}$Institute for theoretical Physics, ETH Zurich, 8093 Zurich, Switzerland}

\date{\today}

\begin{abstract}
Phononic crystals and metamaterials take advantage of pre-designed geometrical structures to sculpt elastic waves, controlling their dispersion using different mechanisms. These mechanisms revolve mostly around Bragg scattering (BS), local resonances (LR) and inertial amplification (IA), which employ ad-hoc, often problem-specific geometries. Here, we use parametrized, spiraling unit cells as building blocks for designing various types of phononic materials. We focus on planar spirals that are easy to fabricate, yet give rise to the desirable complex dynamics. By simple modifications of the spirals, we open full band gaps using BS, LR and IA. Moreover, we alter the underlying unit cell symmetry and lattice vectors, to create wave beaming and topologically protected band gaps, both affecting waves whose wavelength is much larger than the order of periodicity.
\end{abstract}
\pacs{Valid PACS appear here}
\maketitle


Phononic crystals and metamaterials have been used to manipulate waves in a wide frequency spectrum: from heat propagation at very high frequencies, ultrasonic waves at high frequencies, down to audible sound and earthquake excitations at low or very low frequencies\cite{maldovan2013sound}. They generally consist of two- or three-dimensional unit cells arranged in periodic arrays. The advantage of designing building blocks in fundamentally discrete materials -- like phononic materials -- is the ability to engineer their dispersion relation. A common method to control the propagation or reflection of waves in dispersive systems is the opening of frequency band gaps, where waves can't penetrate the material bulk. Moreover, these frequency bands can have unconventional characteristics, for example, directing radially emitted waves to propagate only along a line (wave beaming)\cite{torres1999sonic}, or along the edges of a medium, without being susceptible to imperfections or back scattering (waves with topological protection)\cite{kane2005quantum}.  

Generally, opening a band gap can be accomplished utilizing three different physical phenomena: (i) Bragg scattering (BS), where a periodic medium can inhibit waves whose wavelength is on the order of the mediums' spatial periodicity (i.e., the Bragg limit) through destructive interferences\cite{sigalas1993band,kushwaha1993acoustic}. This is usually achieved by having two materials within the unit cell or a single material with holes. (ii) Local resonances (LR), where wave propagation can be restricted using an inherent resonance in the unit cell\cite{liu2000locally}, decoupling the unit cell size from the wavelength of the attenuated waves and removing the periodicity requirement of the medium\cite{rupin2014experimental}. (iii) Inertial amplification (IA), where a resonator is usually coupled to the unit cell in multiple locations\cite{yilmaz2007phononic}. While this coupling is usually achieved with hinges and rigid connections, multiple connections and chirality are sufficient to observe inertial resonances\cite{bigoni2013elastic}. All these band gap opening mechanisms can be employed to engineer dispersion and manipulate elastic waves for a multitude of applications such as focusing, beaming or shielding and insulation \cite{hussein2014dynamics}.

\begin{figure}[!b]\begin{center}
		\includegraphics[width=8.6cm]{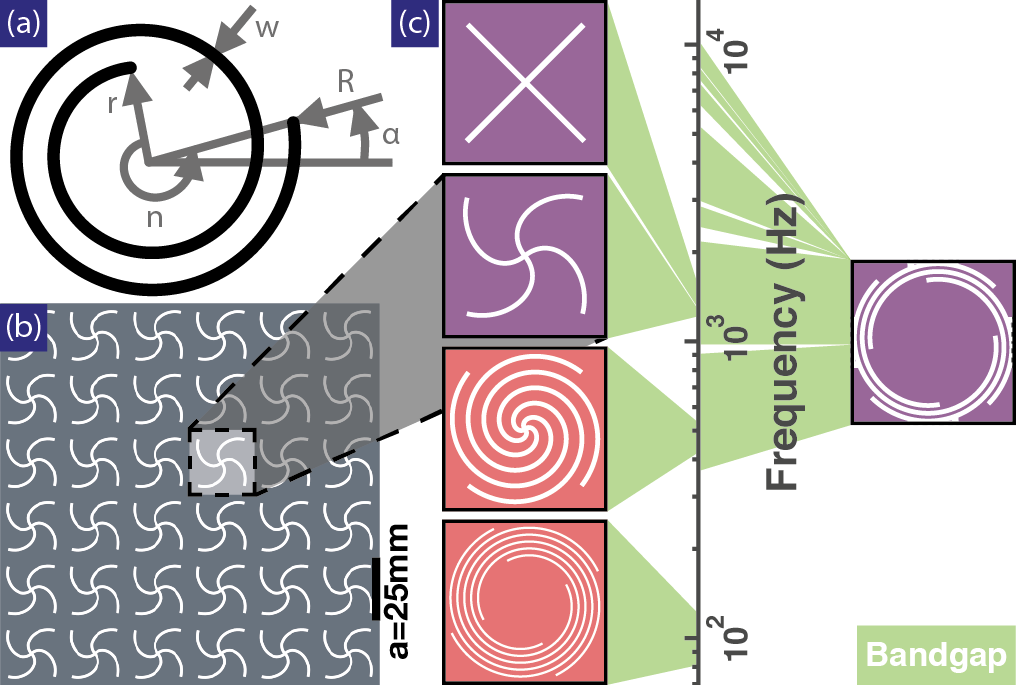}
	\end{center}
	\caption{ {\bf Spiral-based structured plates.\,}{\bf (a)} Schematic of an Archimedean spiral with its parameters. {\bf (b)} Structured plate composed of a periodic array of spirals. White areas indicate the absence of material. {\bf (c)} Varying spiral geometries with flexural band gaps shaded in green. The band gaps span two decades of frequencies, while keeping constant the lattice size, $a$ = 25 mm, thickness, $th$ = 3 mm, density $\rho$ = 1018 Kg/m$^{3}$, Young's Modulus $E$ = 2 GPa and Poisson's ratio = 0.33. Unit cells exhibiting Bragg scattering are colored in purple and local resonances in red.}\end{figure}

The design approaches for inducing each of these band gap mechanisms are vastly different (from material layering and holes for BS, pillar coatings and heavy inclusions for LR to hinged mechanisms for IA)\cite{hussein2014dynamics}. Additionally, the design of specific band properties -- such as doubly negative properties -- adds further complexity\cite{kaina2015negative}. The need for a systematic design methodology -- able to generate a plethora of physical phenomena for a variety of demanding applications -- is apparent. In this article, we present the use of Archimedean spirals as a platform for realizing different phononic metamaterial plates, by simple variations of the spirals' geometrical parameters and symmetry. We show that by varying the spirals's geometry, it is possible to realize BS, LR and IA, with no added masses or hinges. To assess whether the waves affected are longer than the unit cell --i.e. subwavelength-- or not, we calculate the Bragg limit of the metamaterial based on its dispersion relation, rather than referring to the properties of the homogeneous material. We demonstrate various intensities of wave beaming in a homogeneous structured plate, utilizing inertial amplification. Finally, we present a topological metamaterial based on a sub-wavelength resonance induced Dirac cone. These spiral materials can be easily manufactured using additive\cite{bilal2017reprogrammable} or subtractive\cite{bilal2017bistable} manufacturing technologies.

Spirals are common in nature and art\cite{cook1903spirals} and some of their mathematical bases have been known for millennia\cite{heath1897works}. In phononics, spiral-based piezo-patches enabled directional excitation and sensing\cite{yoo2010piezoelectric,baravelli2011double}. Uni-spiral geometries embedded with heavy double-pillared inclusions have been used to show local resonance\cite{zhang2013low} and other chiral structures have already displayed negative refraction\cite{bigoni2013elastic,zhu2014negative} and wave beaming\cite{spadoni2009phononic}. Recently, the phononic tunability of spiral-based metamaterials has been demonstrated\cite{bilal2017bistable,jiang2017dual, bilal2017reprogrammable} and used to realize the first exclusively phononic transistor\cite{bilal2017bistable}. An understanding of the design of spiral-based metamaterials and the extent of physics achievable thereby, however, is still lacking.

The polar representation of a spiral[Fig. 1(a)] is $r(s)= R-(R-r)\,s$, $\phi (s) = 2\pi\, n\, s$, where $r$ is the inside radius, $R$ is the outside radius, $n$ is the number of turns and $s \in [0; 1]$. Such spirals can be repeated periodically in any lattice configuration, producing a plethora of phononic properties. We start with a simple square lattice[Fig. 1(b)] with four concentric spirals (preserving $\mathcal{C}_4$ symmetry\cite{bigoni2013elastic}) as voids within a solid plate. To characterize the dispersion properties of the different unit cells and identify the band gap locations, we model the material using the elastic wave equations for heterogeneous media\cite{graff2012wave} in an infinite lattice and apply Bloch boundary conditions\cite{bloch1929quantenmechanik}. We solve the resulting equation using the finite element method (COMSOL 5.2). The solution is the wavefunction $u (x, \kappa; t) = \tilde{u} (x) \exp (i (\kappa^\intercal x-\omega t))$, where $\tilde u$ is the Bloch displacement vector, $x$ is the position vector, $\kappa$ is the wavenumber, $\omega$ is frequency and $t$ is time. By varying the design parameters of the spirals within a single cell, we open band gaps spanning more than two orders of magnitude in frequency, while keeping constant the plate thickness and lattice spacing[Fig. 1(c), Supp. video 1]. 

\begin{figure}[!b]\begin{center}
		\includegraphics[width=8.6cm]{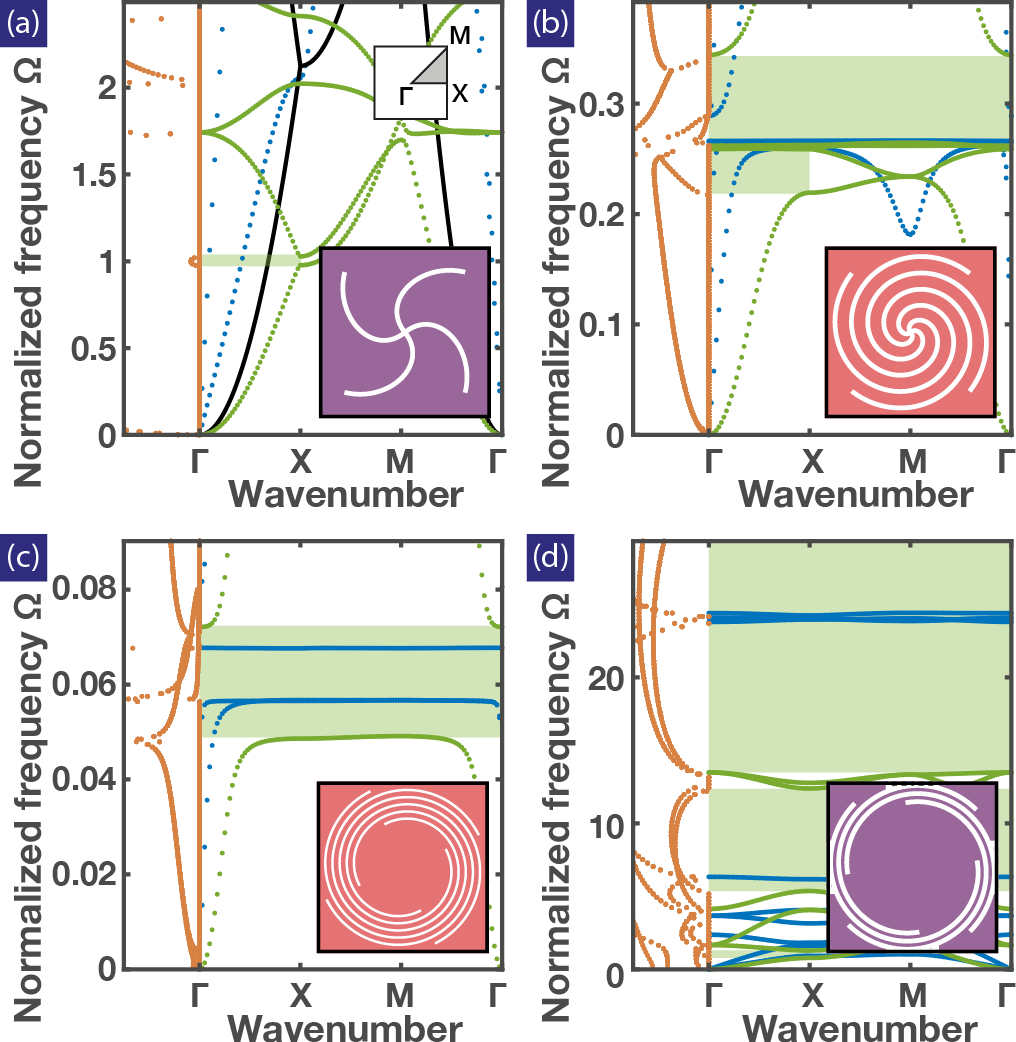}
	\end{center}
	\caption{ {\bf Band gap opening mechanisms.\,}Non-dimensional dispersion curves with both the attenuated (orange) and propagating (blue for in-plane waves and green for out-of-plane waves). The corresponding unit cells are in the inset. Band gaps are shaded in green. {\bf (a)} Bragg scattering, with $r=0$ mm, $n=0.25$, $R=12.8$ mm, $w=0.6$ mm, and $\alpha = 45^\circ$. 
(black line) dispersion for flexural waves in the non-spiral material.
{\bf (b)} Local resonances, with $r=0$ mm, $n=1.25$, $R=12.8$ mm, $w=0.6$ mm, and $\alpha = 45^\circ$. {\bf (c)} Inertial amplification, with $r=6$ mm, $n=1.25$, $R=12.8$ mm, $w=0.6$ mm, and $\alpha = 23^\circ$. {\bf (d)} Bragg scattering, with $r = 10$ mm, $n= 0.65$, $R=13.25$ mm, $w=0.65$ mm, and $\alpha = 37.7{}^\circ$.}
\end{figure}

In all geometries considered, to identify the band gap opening mechanism and determine its position relative to the Bragg scattering limit, we overlay both propagative (i.e., real wavenumbers[Fig. 2(a-d) right]) and attenuating (i.e., imaginary wavenumbers[Fig. 2(a-d) left]) waves at different frequencies. While the Bragg limit in a given homogeneous medium has a fixed value for a given polarization, the introduction of other materials (or voids) within the unit cell alters the Bragg limit significantly[Fig. 2(a) black lines, green dots]. Therefore, we present the frequency dispersion diagrams in nondimensional units  $\Omega = \frac{f}{f_{Bragg}} $, where $f_{Bragg}$ is the computed Bragg limit for each unit cell configuration[Fig. S1], resulting in $\Omega = 1$ at the Bragg limit. By incorporating a minimal spiral void, with $n$ = 0.25, we induce a minute Bragg scattering partial band gap at $\Omega =1$ [Fig. 2(a) green shaded area] with a Bragg-like semicircular band within the attenuation profile[Fig. 2(a) orange dots]. By increasing $n$ to 1.25, four disconnected, concentric beams emerge within the unit cell, acting as local resonators. This opens a subwavelength band gap, with inverted resonance peaks within the attenuation profile[Fig. 2(b)]. By increasing the inner radius of the spiral $r$ from 0 to 6 mm, a single resonant mass emerges at the spiral core connected through four beams creating an inertially amplified unit cell[Fig. 2(c)]. This opens a deep subwavelength band gap with more pronounced inverted peaks within the attenuation profile\cite{frandsen2016inertial,bigoni2013elastic}. By increasing the outer radius $R$ from 12.8 to 13.25 mm the spiral void intersects with the unit cell boundary. This opens multiple Bragg scattering band gaps with semi-circular attenuation profiles[Fig. 2(d)]. Surprisingly, without normalizing the frequency, the first BS gap in [Fig. 2(d)] is similar to that of the LR geometry[Fig. 1(c)]. This reiterates the influence of the considered geometry on its Bragg limit.

To further investigate the effect of the parameters  of the spiral pattern on its phononic band gaps, we vary their values systematically and analyze the resulting wave propagation[Fig. S3]. Results show that the phononic dispersion of a unit cell with a spiral pattern depends mainly on its cross-section, length of the connecting ligaments (the interplay between $w$, $n$, $r$ and $R$) and the spiral core's radius/mass $r$[Fig. S3(c-f)]. Varying the plate's thickness, $th$, allows for tuning the frequencies of flexural waves, without affecting in-plane waves[Fig. S3(f)]. Changing the lattice spacing $a$, while keeping the same spiral pattern, modifies the relative stiffness between the spirals and the plate, hence, reducing the band gap width[Fig. S3(g)]. The orientation of the spirals has little effect on the band gap width[Fig. S3(h)]. 

In addition to the parameters governing the spiral pattern, the underlying unit cell symmetry plays a critical role in its phononic dispersion properties. An asymmetry within the unit cell, resulting in partial band gaps and pass bands, can lead to exotic effects such as directional band gaps and preferential paths for waves within homogeneously structured media (i.e., wave beaming\cite{phani2006wave, zelhofer2017acoustic,ruzzene2005directional, spadoni2009phononic,celli2014low}). Such asymmetry can be introduced by utilizing a different number of spirals within a unit cell. For example, using two concentric spirals in a square lattice results in a $\mathcal{C}_2$ symmetry. In this case, the Brillouin zone should be extended to include half of the first Brillouin zone for a complete dispersion diagram calculation. The resulting beaming effects can be utilized in frequency-dependent waveguiding. 

\begin{figure}[!t]
	\includegraphics[width=8.6cm]{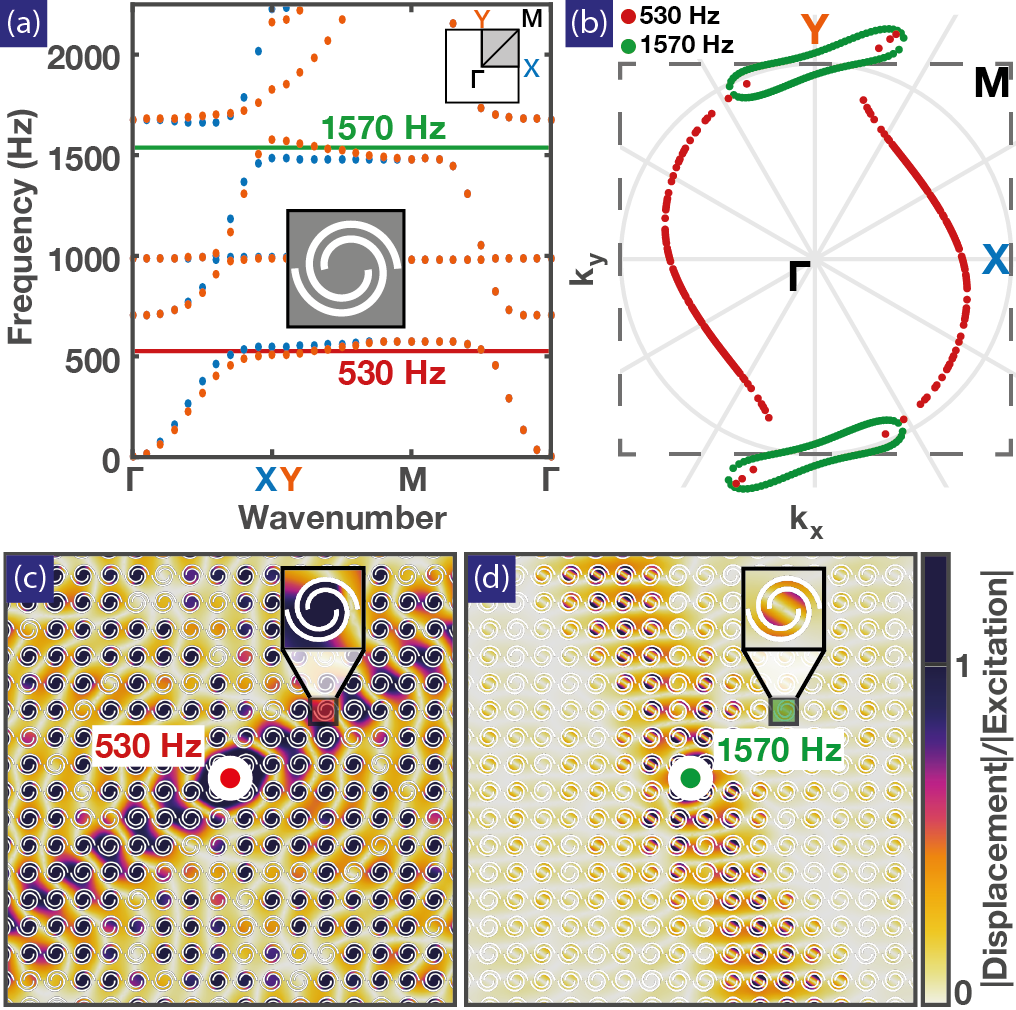}
	\caption{ {\bf Wave beaming.\,(a)} Dispersion relation for the unit cell in the inset ($th=3$ mm, $a=25$ mm, $r=4.3$ mm, $R=11$ mm, $n=1$, $w=1.5$ mm, and $\alpha=0^\circ$). {\bf (b)} Isofrequency plot of the wavenumber for the entire unit cell. The dashed gray lines indicate the unit cell edges in reciprocal space. {\bf (c)} Frequency response of a finite plate, under out-of-plane harmonic excitation at 530 Hz and {\bf (d)} 1570 Hz with absorbing boundary conditions. 
		}\end{figure}

To demonstrate frequency-dependent beaming, we utilize a unit cell created by two concentric spiral voids, resembling a resonator connected within the unit cell in two points (an inertial amplification-like geometry). IA not only decouples the operational frequency from the unit cell size but also relies on resonances that are only excited by specific waves; therefore enabling sub-wavelength beaming[Fig. S2(a,b)]. To identify the frequencies of interest we consider the first quadrant of the $\kappa_x$ and $\kappa_y$ space[Fig. 3(a) inset]. The dispersion relation for flexural waves of the considered unit cell shows a deviation between the dispersion branches along the high symmetry lines $\Gamma$-X (blue) and $\Gamma$-Y (orange)[Fig. 3(a)]. We harness this deviation, manifested in the same unit cell, to show two wave beaming at two different frequencies: (i) one in a pass band, at $f_1= 530$ Hz (red line), and (ii) one in a stop band, at $f_2 = 1570$ Hz (green line). A partial band gap exists at $f_1$ in the $\Gamma$-Y direction, with a pass band at $f_1$ elsewhere. While a partial pass band exists at $f_2$ in the $\Gamma$-Y direction, with a band gap at $f_2$ elsewhere. The projection of the Eigen-solutions on the full reciprocal space[i.e., the iso-frequency contours, Fig. 3(b)] provides a complete picture. At $f_1$ (red) there exists no propagation along the $\Gamma$-Y direction, with a strong preference for the waves to propagate along the diagonal of the first quadrant of the reciprocal space. At $f_2$ (green), the propagation is confined only along the $\Gamma$-Y direction. The response of a finite sample with absorbing boundary conditions[Fig. 3(c-d)] --driven harmonically out-of-plane in the center, at $f_1$ and $f_2$ separately-- indicates a response similar to the infinite medium predictions. Within a pass band, at $f_1$, there exists a strong confinement of waves along a line, while waves still propagate radially in the plate. Within a band gap, at $f_2$, the wave propagation is also confined along a line, however, with no propagation is allowed anywhere else in the plate. Tuning the direction and intensity of the propagating waves can be achieved by changing the orientation of the spirals within the unit cell, as well as by selecting different operating frequencies[Fig. S4].

\begin{figure*}[!t]
		\begin{center}
		\includegraphics{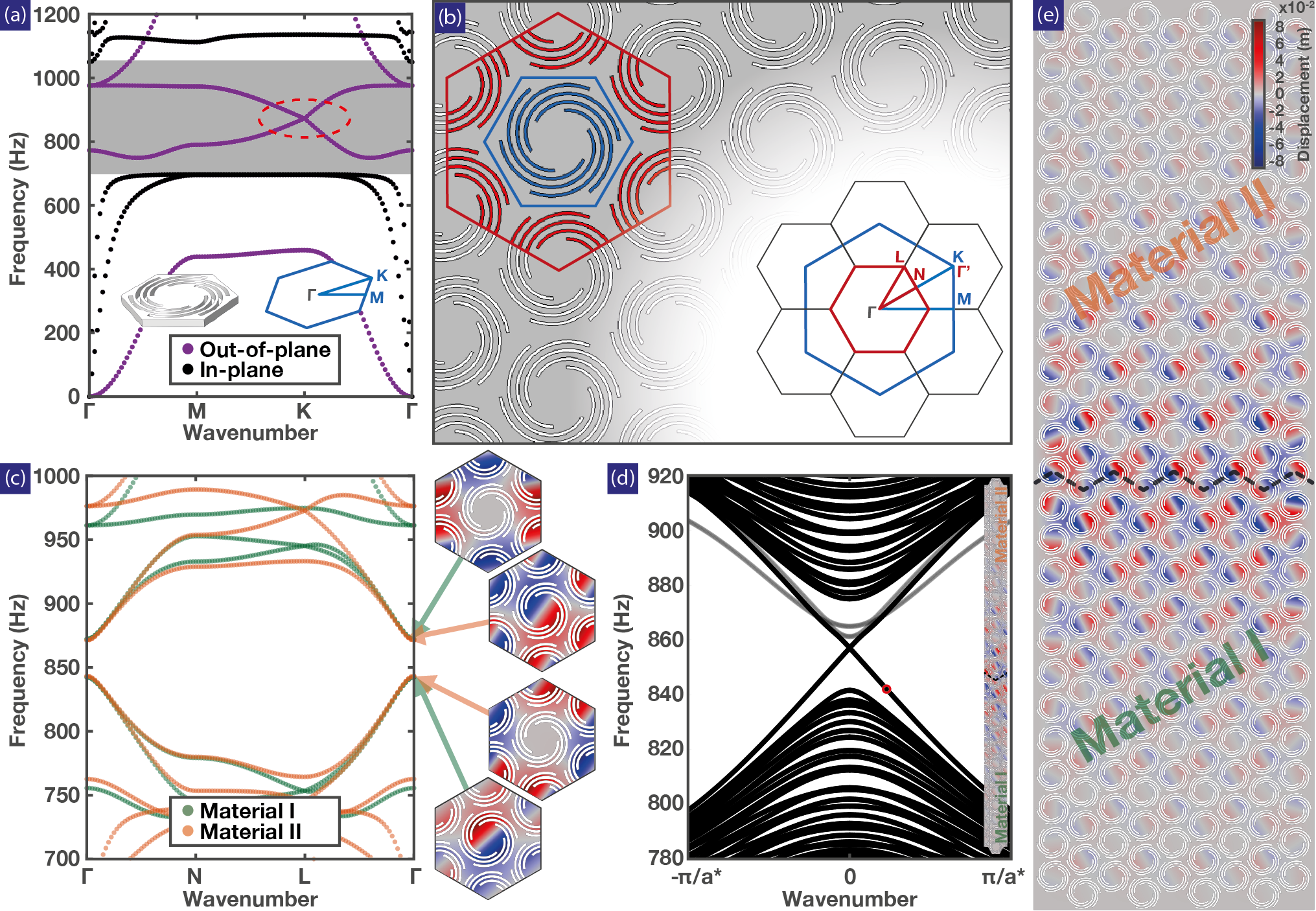}
		\end{center}
\caption{ {\bf Topological insulators.\,}{\bf (a)} Dispersion curves of a hexagonal unit cell and its IBZ ($a$ = 25mm, $th = 2 $ mm, $r = 6.6$ mm, $R = 12$ mm, $n = 0.5$, $w = 1$ mm). In-plane band gap highlighted in gray and a subwavelength Dirac cone marked in red. {\bf (b)} Left: super cell that includes the original hexagonal unit cell (blue) with additional neighboring spirals (red hexagon) in real space. Right: the reduced Brillouin zone ($\Gamma$-N-L-$\Gamma$) for the modified spiral patterns in reciprocal space. {\bf (c)} Two dispersion curves of two super cells with small variations between the spirals at the core vs at the boundaries (supercell I: $n_{\mathrm{blue}}=1.02\, n$) and (supercell II: $n_{\mathrm{blue}}=1.03\, n$, $w_{\mathrm{blue}}= w_{\mathrm{red}} = 1.068\, w$). The insets show the mode shapes of each unit cell at the edge of the topological gaps for super cells. {\bf (d)} Dispersion curves for a quasi-finite (periodic in one direction and finite with 27 spiral unit cells in the other) line of material I and II. In addition to trivial edge bands (gray), the previously observed band gap is closed by two topologically protected counter-propagating modes. The mode shape corresponding to the circled dispersion point (red), with the interface between the materials indicated by a black dashed line. {\bf (e)} The mode shape at frequency $f$ = 857 Hz for a finite sample composed of materials I and II.}\end{figure*}

The lattice vectors that govern the spacial packing of the spiral patterns can have a significant influence on the nature of the band gaps. 
This is particularly relevant for band gaps induced by Bragg scattering, since for band gaps resulting from resonances (LR and IA), periodicity is less important. However, some of the emerging phonon physics can benefit greatly from both periodicity and resonances. For example, band gaps that are protected by topology within a periodic lattice can prohibit waves from penetrating through the material bulk. These topological materials, however, support the propagation of scattering immune waves along their boundaries, regardless of the presence of defects\cite{susstrunk2015observation, mousavi2015topologically, brendel2017snowflake, pal2017edge}. Opening such topological gaps using resonances enables the realization of subwavelength topological insulators, which decouple the lattice spacing from the operating frequency\cite{yves2017crystalline,yves2017topological}.

One avenue to design a subwavelength topological insulator is the opening of a band gap within a degenerate (duplicate) Dirac cone. Hence, we start by creating a single Dirac cone for flexural waves --which is known to exist in a hexagonal packing-- at a subwavelength frequency, $f$. We utilize a unit cell with six concentric spirals, resembling a resonator connected at six points to the unit cell (preserving $\mathcal{C}_6$ symmetry). As concluded from Fig. S3(f), the thickness of the plate can be a design parameter to independently control the band gap frequencies of flexural and in-plane waves. Therefore, we choose the unit cell thickness, $th=$ 2 mm, to ensure a full band-gap for in-plane waves in the vicinity of $f$ (873 Hz)[Fig. 4(a)]. It is worth noting that the Bragg limit for the utilized design is 1661 Hz[Fig. S4(c)]. 

In order to duplicate the Dirac cone, we consider a super cell (red) encompassing the original unit cell (green) in addition to one-third of each of the neighboring unit cells[Fig. 4(b) left]\cite{wu2015scheme,chaunsali2017subwavelength}. Such an artificial enlargement of the unit cell boundaries, results in doubling the branches within the dispersion curves through the folding of the cell's irreducible Brillouin zone (IBZ)[Fig. 4(b) right]. By modifying the spiral pattern of the original (green) and the neighboring (red) unit cells --breaking its translation symmetry, while preserving its $\mathcal{C}_6$ one-- the supercell becomes the smallest unit cell. The resulting unit cell manifests a topologically protected band gap at the $\Gamma$ point\cite{brendel2017snowflake, kaina2015negative}, as a degenerate, yet inverted, mode exists at both edges of the gap[Fig. 4(c)]. By tiling two variations of the unit cell (as material I and II), a topological wave guide emerges at the interface\cite{brendel2017snowflake}. To demonstrate the existence of a topological state at the interface between the two materials, we first calculate the dispersion curves of a quasi-finite sample (periodic in one direction and finite along the other) composed of the two materials[Fig. 4(d)]. The two bands at the edge of the gap connect through two straight lines crossing at the $\Gamma$ point, with a mode shape showing clear localization of the wave at the interface[Fig. 4(d,inset)]. Then, we simulate a plate composed of the two materials, in Fig. 4(e), to study the steady state response of waves confined along the interface between the two materials (highlighted with a dashed black line) at $f$.

Through a systematic analysis, we show that the introduction of spiral patterns can greatly alter the dispersion characteristics of phononic media. By introducing these spiral patterns as voids within solid plates, we realize different phononic band gaps: Bragg scattering, local resonance and inertial amplification. We capitalize on the internal unit cell symmetry, to create partial propagating bands within full band gaps and demonstrate two classes of subwavelength wave beaming, one within a propagating frequency and one within an attenuating one. Moreover, we alter the underlying lattice symmetry to create subwavelength induced topologically protected band gaps. Our demonstrations show the potential of coupling various wave phenomena with a unified, easy-to-design and fabricate building blocks, enabling the design of novel phononic material plates for different applications. The presented advances highlight the versatility of spirals-based phononic systems.

\setcounter{figure}{0}
\renewcommand{\thefigure}{S\arabic{figure}}
\section{Appendix}

\subsection{(A) Bragg limit approximation}
One way a periodic medium can interact destructively with waves is by the unit cell size being a positive integer multiple of the wavelength. The lowest frequency around which this happens is referred to as the Bragg limit. For a given homogeneous material, the speed of sound is usually fixed. Therefore, by having wave velocity and the order of spacial periodicity (i.e., the unit cell size),  the Bragg limit can be calculated. However, introducing cuts within the unit cell reduces the effective properties of the material, thus changing its dynamical characteristics. In addition, for flexural waves in plates, the wave velocity is not constant. The dispersion is rather parabolic [Fig. S1(a)], which adds complexity in determining the limit at which periodicity can no longer affect waves. Such parabolic nature shows an increase in the slope of the dispersion branch (i.e., phase velocity) as the wavenumber increases. By identifying the maximum of this phase velocity, we approximate the Bragg limit to be at the edge of the first Brillouin zone (i.e., the high-symmetry point X). In order to determine that phase velocity maximum, we apply a standard finite difference scheme to the dispersion bands. We utilize this scheme to determine the Bragg-limit for the designs presented in figure 2. We numerically determine the point of maximum phase velocity (black) for flexural waves in the $\Gamma-X$ direction. We use the tangent of the dispersion branch at these points (black dots) to project the Bragg-limit frequency. The frequencies susceptible to periodicity (i.e., higher than the  Bragg limit) are highlighted with a gray dashed background.


\begin{figure}
	\begin{center}
		\includegraphics[width=7.95cm]{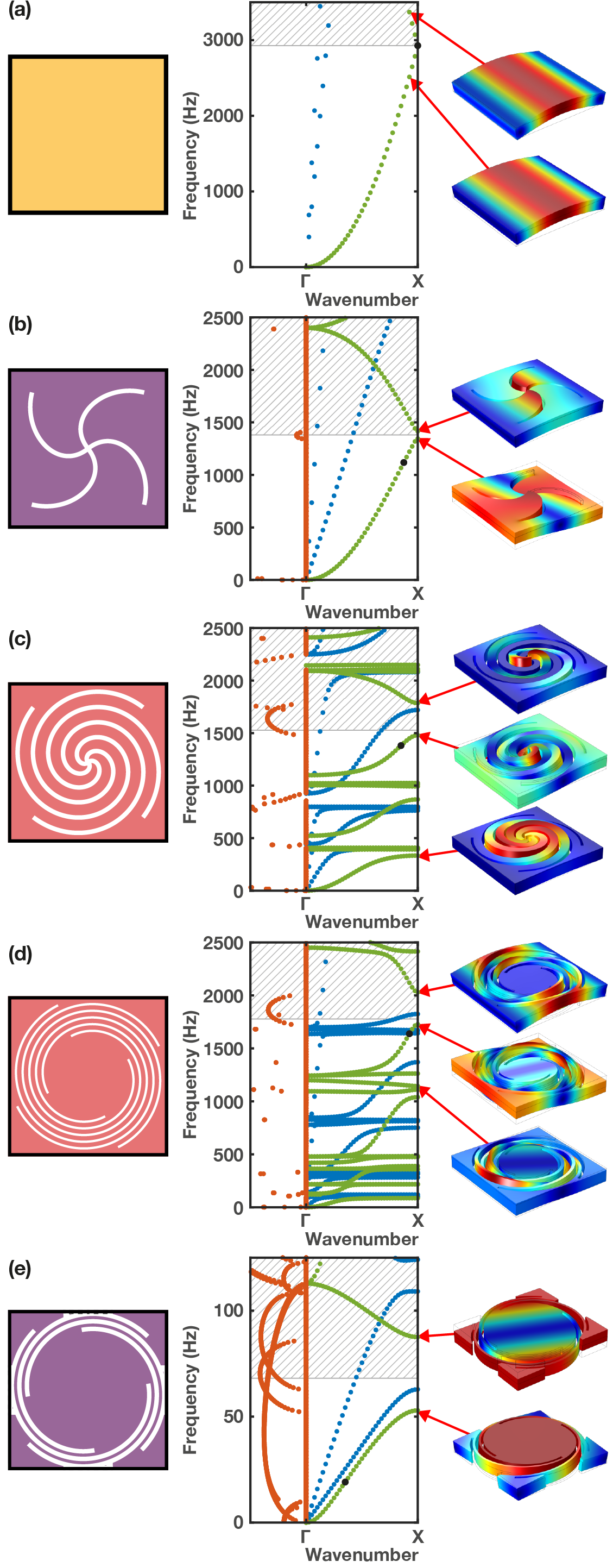}
	\end{center}
	\caption{ {\bf Bragg limit computation.\,}(a-e) Dispersion curves for both attenuated (orange, b-e) and propagating waves (in-plane: blue, and out-of-plane: green). The effective wave velocity is tangential in the point of maximum phase velocity (black). The cells' Bragg limit is where the tangent meets the $\Gamma$-point, frequencies beyond this limit are indicated by a gray background. The respective unit cell is inset on the left.}\end{figure} 

By applying the method outlined above, we compute the Bragg limit for the unit cells of figure 2 in the main text. We start by calculating the Bragg limit for the homogeneous material used throughout the paper (VeroWhite: density $\rho$ = 1018 Kg/m$^{3}$, Young's Modulus $E$ = 2 GPa and Poisson's ratio = 0.33) [Fig. S1(a)]. The maximum phase velocity coincides with the $X$-point and thereby the Bragg limit, which is well established for homogeneous media in the literature  \cite{hussein2014dynamics}. Each of the mode shapes at the Bragg limit of ~3000 Hz shows a wave on the order of the unit cell size [Fig. S1(a)].  Once an extra feature is added to the unit cell (e.g., a spiral cut), the Bragg limit changes. For example, by introducing a simple spiral cut to the unit cell and inducing a minute band gap, the Bragg limit shifts below 1500 Hz. To confirm that the induced gap is of the Bragg type, we plot the corresponding imaginary solution of the dispersion relation (orange) [Fig. S1(b)]. By changing the pattern of the spiral cut through the plate [Fig. S1(c-d)], the position of the Bragg scattering induced band gaps changes, revolving around 1500 Hz. However, this number is reduced by two orders of magnitude, when the spiral patten touches the boundaries of the unit cell, opening a Bragg-scattering induced gap at 50 Hz [Fig. S1(e)]. In all of the plotted insets, the mode shapes below the Bragg limit show strongly localized vibrations[Fig. S1(c,d)]. In conclusion, the Bragg limit is significantly altered by the introduction of the geometry. The Bragg limit is $\sim1/2$ that of the homogeneous medium for the geometries in [Fig. S1(b-d)] and $\sim1/40$ for the last geometry[Fig. S1(e)]. Referring to the Bragg limit of the homogeneous material is thus inadequate to identify whether the waves \emph{within} the metamaterial are smaller than its' unit cell size, i.e. subwavelength.

Following the same methodology, we calculate the Bragg-scattering limit for both geometries presented in beaming and topological wave protection. The Bragg limit in the beaming case is 2298 Hz in $\Gamma-X$ and 1841 Hz in $\Gamma-Y$ directions[Fig. S2(a,b)]. We also apply the same method to identify the Bragg limit of the topologically protected material[Fig. 4(a)]. We consider the simple unit cell to verify that the topological insulator is based on subwavelength resonances. For bending bands along $\Gamma-M$[Fig. S2(c)], the Bragg limit for the initial unit cell with the Dirac cone is 1661 Hz. The Dirac cone is centered around 873 Hz and thus subwavelength. Therefore, in both cases (beaming and topological insulation) the design successfully decouples the unit cell size from the wavelength.


\begin{figure}
	\begin{center}
		\includegraphics[width=8.6cm]{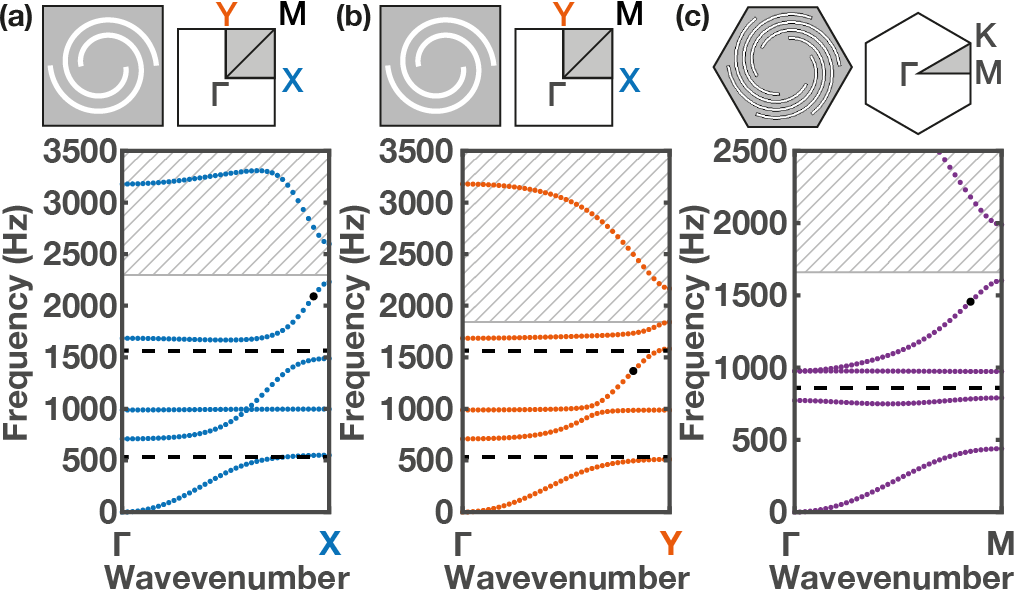}
	\end{center}
	\caption{{\bf Subwavelength Beaming and Dirac cones.\,}Dispersion relation for out-of-plane waves of the unit cells (top). The maximum effective wave velocity is indicated by a black dot, frequencies that exceed the Bragg limit are indicated by a gray dashed background.  The frequencies of operation for both demonstrations of beaming and topological insulator are indicated by black dashed lines. For the beaming unit cell in {\bf(a)} $\Gamma-X$ direction, and {\bf(b)} in $\Gamma-Y$ direction. {\bf(c)} for the hexagonal TI cell in $\Gamma-M$ direction}\end{figure}

\subsection{(B) Design of spiral geometries}
The parameter space for spiral-based phononic metamaterial is vast, with multiple design dimensions. To capture the influence of each parameter on the resulting pattern, we systematically vary some of these parameters and record the evolution of the first three flexural band gaps [Fig. S3].  All parameters are kept constant except for the one being analyzed. The reference parameters are: the unit cell size $a = 25$ mm, the plate thickness $th = 3$ mm, the spiral width $w = 0.6$ mm, the inside radius $r = 6$ mm, the outside radius $R = 12.84$ mm, the number of turns $n = 1.25$ and the rotation $\alpha =23^\circ$ [Fig. S3(a,b)]. These reference parameters result in a central mass connected to the plate frame by thin spiraling ligaments. Changing the dimensions of these ligaments, or its connected central mass, influences the frequency response of the metamaterial. For instance, varying the number of turns, while keeping both the inner and outer radius constant, affects both the length and the width of the connecting ligaments. Therefore, as the number of turns increases, the band gap frequencies decrease significantly [Fig. S3(c)]. Similarly, increasing either the inner radius [Fig. S3(d)] or the cutting width of the spirals [Fig. S3(e)] has a similar effect (i.e., lowers the frequencies) on the band gaps, as it increases the central mass or reduces the width of the connecting ligaments. By increasing the plate thickness, the band gap frequencies increase at the beginning, for very small thickness, but eventually saturates. Since variations in the ligaments' thickness do not have a profound effect on in-plane waves, changing the thickness is an ideal tuning parameter to independently change the position of flexural band gaps[Fig. S3(f)]. Both the lattice constant [Fig. S3(g)] and the spiral orientation angle [Fig. S3(h)] seem to have little to no influence on the position of the band gaps.


\begin{figure}
	\begin{center}
		\includegraphics[width=8.6cm]{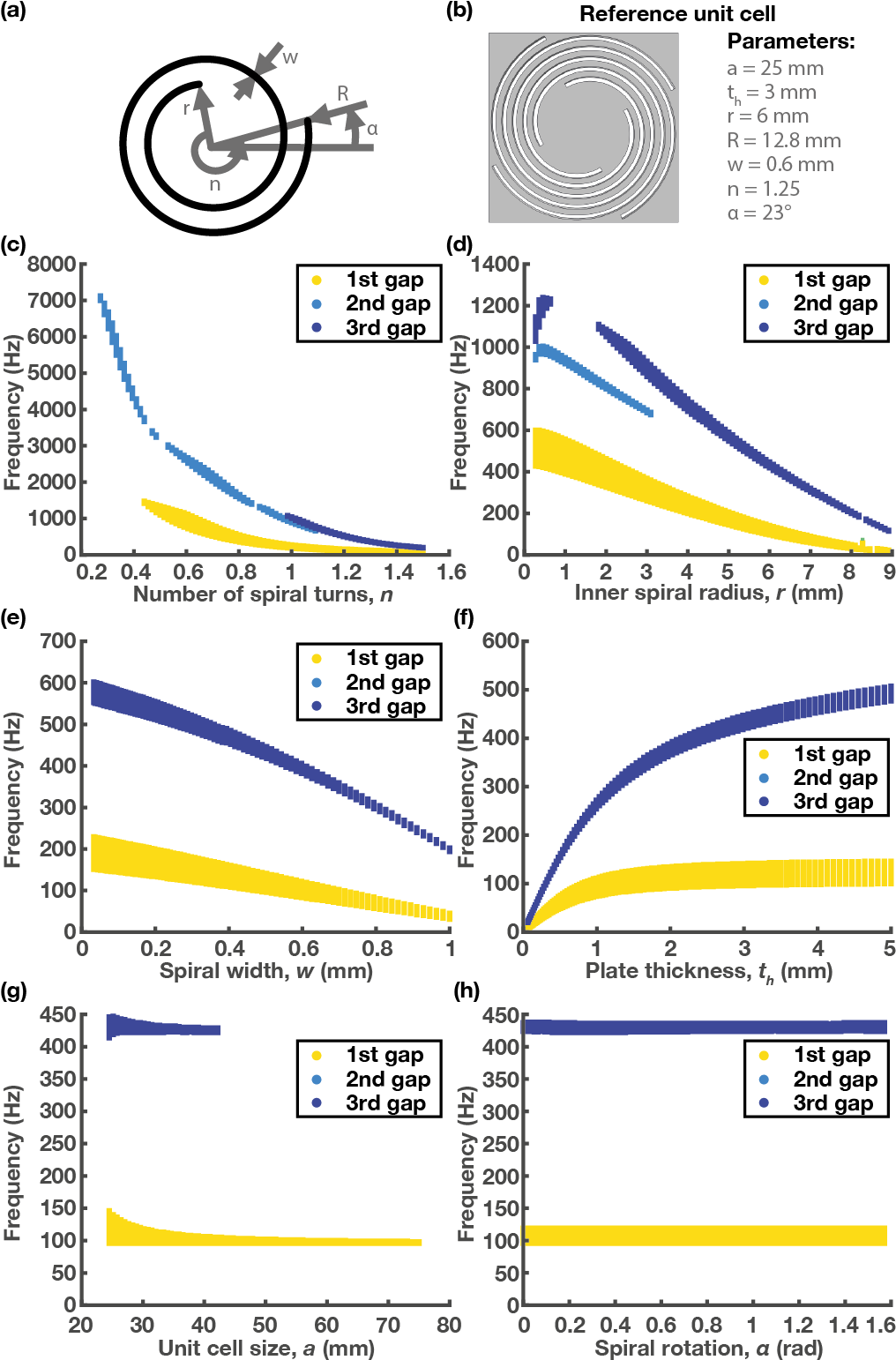}
	\end{center}
	\caption{{\bf Parameter influence.\,} (a) Geometric parameters of an Archimedean spiral and (b) the unit cell used as a reference for the parameter analysis. (c-h) Out-of plane band gaps of the inertially amplified unit cell for various parameters. We vary (c) the inside radius of the spirals, $r$, (d) the number of turns, $n$, (e) the spiral width, $w$, (f) the plate thickness, $th$, (g) the unit cell size, $a$, and (h) the orientation, $\alpha$ of the spirals.}\end{figure}


\subsection{(C) Wave beaming as a function of the spiral orientation angle}
As presented in figure 3 in the main text, spiral metamaterials can present desirable wave beaming within partial pass bands and band gaps. The  wave beaming patterns can be tuned by varying the excitation frequency. for instance, at orientation angle $\alpha = -30^\circ$, the wave pattern is fundamentally different at $f$ = 490, 510, 560 Hz [Fig. S4(c)]. At $f$ = 490 Hz, the frequency contours are connected with no strong beaming present with a mostly radial propagation pattern. At $f$ = 510 Hz, the frequency contours start to separate into two distinct domains resembling radial propagation with absence of propagative waves in the vertical direction. At $f$ = 560 Hz, the frequency contours separate further into four islands, presenting two beaming lines. For the same orientation angle $\alpha$, at the vicinity of the second band gap, strong dependency on frequency is also evident [Fig. S4(b)]. For example, at $f$ = 1520 Hz, the frequency contours are connected, which translates to radial propagation with a slight preference for waves to propagate along the vertical and horizontal directions. The separation of these connected contours takes place around $f$ = 1540 Hz. More interestingly, between $f$ = 1560 and 1600 Hz, the frequency contours take the shape of two isolated islands. At $f$ = 1560 Hz, a single beaming line in the vertical direction is present. As the frequency increases, the beaming orientation exhibits a clockwise shift from the vertical direction.

Moreover, the properties of the beaming can be controlled further by tuning the orientation $\alpha$ of the spirals spanning a wide range of frequencies.
For example, the isofrequency contours for rotated identical spiral-cuts are plotted in [Fig. S4]. The rotation angles vary from $-30^{\circ}$ to 45$^{\circ}$. It is evident that the orientation of the spiral influences the pattern of wave beaming, even at the same frequency. For example, in the vicinity of the first band gap, waves excited at $f$ = 560 Hz will have two different propagation patterns at orientation angle 0 and 30 [Fig. S4(c)]. At $\alpha$ = 0$^{\circ}$, the wave will propagate in two perpendicular directions radiating from the four corners of the unit cell. While at $\alpha$ = 30$^\circ$ , the wave will propagate along a single direction passing through the center of the unit cell. Similarly, in the vicinity of the second band gap  [Fig. S4(b)] at $f$ = 1600 Hz and orientation angle $-20^\circ$ and 30$^\circ$. When $\alpha = -20^\circ$ the beaming occurs along a straight line, while $\alpha = -20^\circ$ causes the wave to radiate from the corners of the unit cell in two perpendicular directions.

\begin{figure*}
	\begin{center}
		\includegraphics[scale = 1]{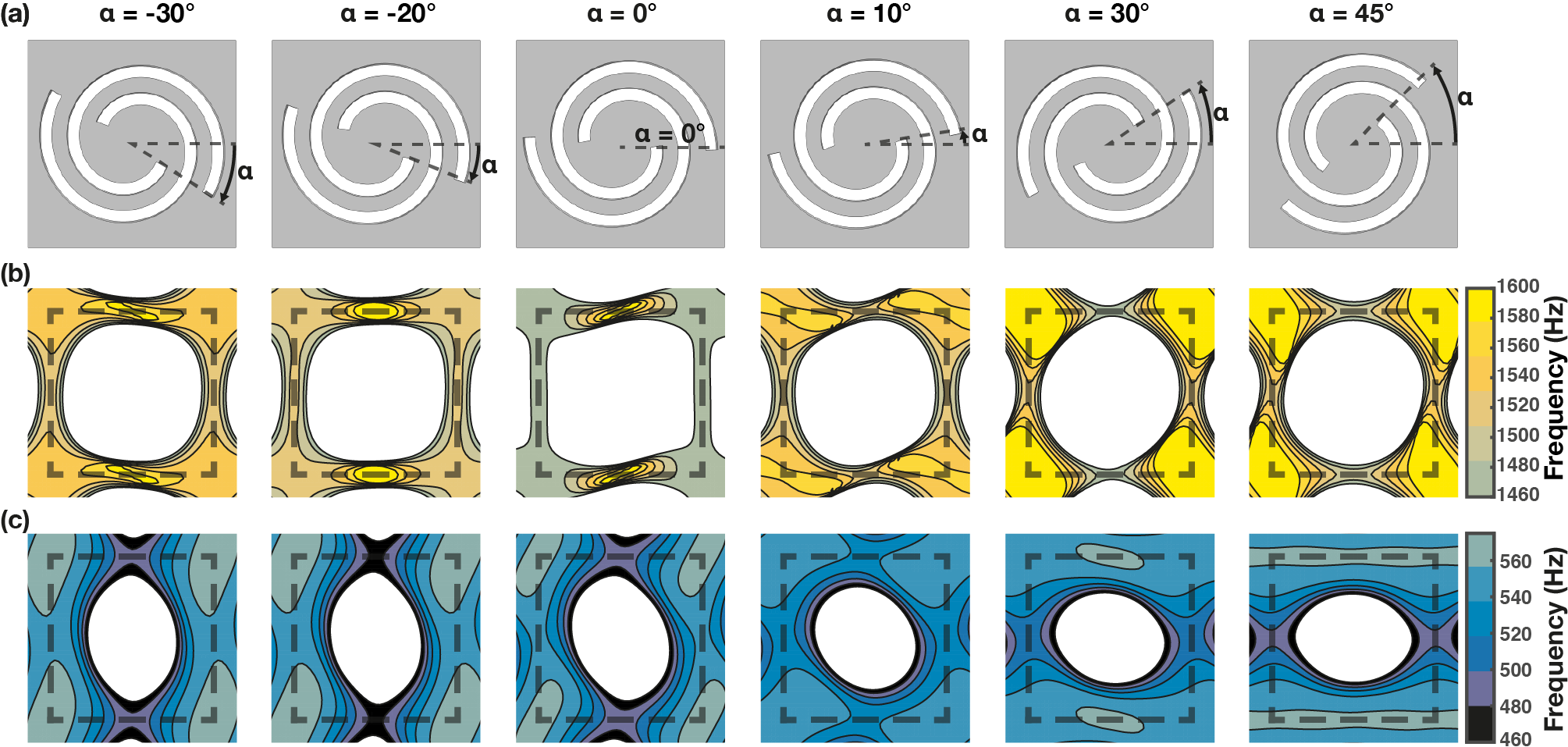}
	\end{center}
	\caption{{\bf Beaming parametric tunability.\,}Isofrequency plot for out-of-plane waves of the unit cell presented in Fig. 3(a) with changing orientations of the spiral (columns), for the two frequency ranges (rows, colorbar). The dashed square indicates the first Brillouin zone of the unit cell.}	
\end{figure*}



\end{document}